\documentclass[amsmath,amssymb,twocolumn,showpacs]{revtex4}
\usepackage{graphicx}
\usepackage{color}
\usepackage{amsmath}
\usepackage{amssymb}
\DeclareMathSymbol{\lang}{\mathord}{symbols}{"68}
\DeclareMathSymbol{\rang}{\mathord}{symbols}{"69}
\DeclareMathSymbol{\openbra}{\mathord}{symbols}{"68}
\DeclareMathSymbol{\closeket}{\mathord}{symbols}{"69}

\newcommand{\aver}[1]{{\lang #1 \rang}}
\newcommand{\be}{\begin{equation}}
\newcommand{\ee}{\end{equation}}
\begin{document}
\title{ Velocity selective trapping of atoms in a frequency-modulated laser field}
\author{V. Yu. Argonov}
\affiliation{argonov@poi.dvo.ru, +7(4232)686592,
Laboratory of Geophysical Hydrodynamics,
Pacific Oceanological Institute of the Russian Academy of Sciences, 
43, Baltiiskaya Street, Vladivostok, Russia 690041}

\begin{abstract}The wave function of a moderately cold atom in a stationary 
near-resonant standing light wave delocalizes very fast due to wave packet
splitting. However, we show that frequency modulation of the field 
may suppress packet splitting for some atoms having specific
 velocities in a narrow range. These atoms remain localized in a small space
for a long time. We propose that in a real experiment with cold atomic gas this
 effect may decrease the velocity distribution of atoms (the field traps the atoms with 
such specific velocities while all other atoms leave the field).
                                  

\end{abstract}
\maketitle

\section{Introduction}

 Laser cooling and trapping of atoms and ions is a rapidly developing
field of modern physics. Cold particles in a laser field are a common physical
substrate used in numerous fundamental and applied issues such as
 Bose-Einstein condensates, quantum chaos, single-atom laser,
 quantum computer,
 etc. A significant number of methods of atomic cooling in a laser field were
 developed in the recent decades (the Doppler cooling \cite{Hansch,Wineland},
the Sisyphus cooling \cite{dalibard,bezverbny}, the velocity selective coherent population
 trapping (VSCPT) \cite{Aspect}, dynamical localization and trapping \cite{RS2011}, etc. \cite{phillips}). Modern sophisticated
 methods provide temperatures of the order of 100 picokelvin \cite{finns}. 

In this paper we suggest a method of coherent laser cooling in the absence
of spontaneous emission. When an atom moves in a near-resonant standing light wave, 
two periodic optical potentials form in the space \cite{Kaz}. When the atom
 crosses a standing wave node, it may undergo the Landau-Zener (LZ) transition
 between these two potentials. Such transitions cause  splitting of the wave 
packets \cite{pra} and rapid delocalization of the wave function
 \cite{JETPL2009}. In this paper we show
 that frequency modulation of the field may suppress the splitting of wave
 packets for atoms that have velocities
 in the specific narrow range (determined by the field modulation parameters). We suppose
that in a real
experiment, this may significantly decrease the energy distribution of
 moderately cold atoms. This method does not pretend to establish any
 temperature record, however, it might be
 useful in some experiments due to its conceptual simplicity.

The ideology of this method is similar to VSCPT and dynamical trapping in some aspects. 
The analogy with VSCPT is rather gentle. Both VSCPT and our method do not cool
 initially "hot" atoms, they only trap the atoms that already have specific 
velocities.
 However, in our method, this velocity is non-zero, and the particular trapping mechanism
 differs from VSCPT radically. Our method is not based on "the dark states". 
It is based on the synchronization between the LZ transitions
 and the field modulation. The analogy with dynamical localization and trapping is more deep. Dynamics of
 cold atoms 
in a periodically modulated (and kicked) standing wave has been studied both
theoretically and experimentally for 20 years by the groups Raizen and Zoller
\cite{RS2011, GSZ92, MR94}. A lot of effects related to dymanical chaos and
quantum-classical correspondence were reported. In particular, it was shown that
in a modulated field, some atoms with special initial positions and momentums can be dynamically trapped
(without obvious energy conditions for such trapping). In terms of dynamical system theory, 
these atoms are trapped in a resonance islands embedded in a chaotic sea (in a phase space) \cite{RS2011}. 
In our study, resonance between field modulation and atomic mechanical oscillations plays similar role. 
However, cited works describe semiclassical atomic motion far from
atom-field resonance. Therefore, there is only one effective optical potential (with modulated amplitude). 
In our study, there are two optical potentials and LZ tunnelings between them. 
This physical situation differs significantly.  

In our study the reported effect was initially proposed theoretically (semiclassical model)
 and then confirmed numerically (purely quantum model). However, we have organized this paper in an alternative order for better  understanding. First, we 
demonstrate the numerical manifestations of the velocity selective trapping, and then
explain the effect theoretically.

\section{Equations of motion}

Let us consider a two-level atom (with the transition frequency
 $\omega_a$ and mass $m_a$) moving in a strong standing laser wave
 with the modulated frequency $\omega_f[t]$. Let us assume that the depth of
 modulation is neglible in comparison with the
average value of frequency $\aver{\omega_f[t]}$ (but not with the detuning
 $\omega_f[t]-\omega_a$), 
so we can consider the corresponding wave vector $k_f$ a constant. In 
absence of spontaneous emission (the atomic excited state must
have long lifetime, or some experimental methods must be used to suppress
 the decoherence) the atomic motion may be described by the Hamiltonian    
\begin{equation}
\begin{aligned}
\hat H=& \frac{\hat P^2}{2m_a}+\frac{1}{2}\hbar(\omega_a-\omega_f[t])\hat\sigma_z-\hbar \Omega\left(\hat\sigma_-+\hat\sigma_+\right)\cos{k_f\hat X},
\label{Jaynes-Cum}
\end{aligned}
\end{equation}
where $\hat\sigma_{\pm, z}$ are the operators of transitions between the atomic
excited and ground 
 states (the Pauli matrices), $\hat X$ and $\hat P$ are the operators of the atomic
 coordinate and momentum, and $\Omega$ is the Rabi frequency. This Hamiltonian was
 used in \cite{pra,JETPL2009,PLA2011}, though for a constant field without 
modulation.

Let us use the following dimensionless normalized
 quantities: momentum  $p\equiv P/\hbar k_f$, time $\tau\equiv \Omega t$,
 position $x\equiv k_f X$,  mass  $m\equiv   m_a\Omega/\hbar k_f^2$ and
 detuning $\Delta[\tau] \equiv(\omega_f[\tau]-\omega_a)/\Omega$. 
Let us suppose that the field modulation is harmonic,
\be 
\Delta[\tau]=\Delta_0+\Delta_1\cos[\zeta \tau+\phi],
\label{mod} 
\ee
and apply 
the following conditions: $\zeta\ll 1$,
 $\Delta_0\lesssim\Delta_1\ll 1$. Using these approximations  
we obtain the
equations for the probability amplitudes to find an atom with the
 normalized momentum  $p$ in the excited or ground state, $a[p,\tau]$ and
 $b[p,\tau]$, correspondently:
\be
\begin{aligned}
i\dot a[p,\tau]&=\left(\frac{p^2}{2m}-\frac{\Delta[\tau]}{2}\right)a[p]-\frac{1}{2}(b[p-1]+b[p+1]),\\
i\dot b[p,\tau]&=\left(\frac{p^2}{2m}+\frac{\Delta[\tau]}{2}\right)b[p]-\frac{1}{2}(a[p-1]+a[p+1])\\
\end{aligned}
\label{qu}
\ee
Here the dot designates the differentiation with respect to $\tau$.
 For every value of $p$, there is its own pair (\ref{qu}).
                            
Let us choose the
 values of the parameters and initial conditions in order to perform the numerical simulation.
The average initial atomic momentum $\aver{p[0]}$ will be a variable condition
for the purpose of this paper. All other conditions will be fixed: normalized mass
 $m=10^{5}$ (by order of magnitude this
 corresponds to the experiments with Cs \cite{Amm98} and
 Rb  \cite{Hens2003} atoms, but for a stronger field $\Omega\sim 10^{9-10} $\,Hz),
field parameters $\Delta_0=-0.02$, $\Delta_1=0.047$, $\zeta=0.00508$,
 $\phi=0$, and the initial form of wave packet 
\be
\begin{aligned}
a[p,0]=b[p,0]=\frac{1}{\sqrt{2\sigma_p[0]\sqrt{2\pi}}}\exp\left[\frac{-(p-\aver{p[0]})^2}{4\sigma^2_p[0]}\right].
\label{i}
\end{aligned}
\ee
Therefore, the initial wave packet has a Gaussian form with $\aver{x[0]}=0$ and the
 initial probability to find the atom in the excited state 0.5. Here $\sigma_p$ is the standard deviation of the atomic momentum
 (equal to the half-width of the packet by order of magnitude).  At $\tau=0$ we
 fix it by the value of $\sigma_p[0]=5\sqrt{2}$. Therefore, in accordance with
 the Heisenberg relation, the standard deviation of
 the initial coordinate is $\sigma_x[0]=1/(2\sigma_p[0])=0.1/\sqrt{2}$ (it is
 much less than the normalized optical wavelength $2\pi$).

In numerical experiments, we use these initial conditions to simulate the system of 8000 equations (\ref{qu}) with
 $-1000\leq p\leq 1000$. For larger values of $|p|$, we put
 $a[p,\tau]=b[p,\tau]=0$  due to the energy restrictions. Obtaining the solution
in the momentum space we perform the Fourier transform and get the wave function
 in the coordinate space in the range of $-4\pi<x\leq 4\pi$ (see figures in the next section).

\section{Numerical results}

In \cite{pra,JETPL2009}, the atomic motion was studied in absence of field
 modulation. The following basic modes of motion were reported. 

At $\Delta=0$ and $|\Delta| \gtrsim 1 $ the atomic motion is simple. Atoms
 move in constant spatially periodic potentials. Slow atoms are trapped in
 potential wells and fast atoms move ballistically through the wave. 

\begin{figure}[htb]
\begin{center}
\includegraphics[width=0.48\textwidth,clip]{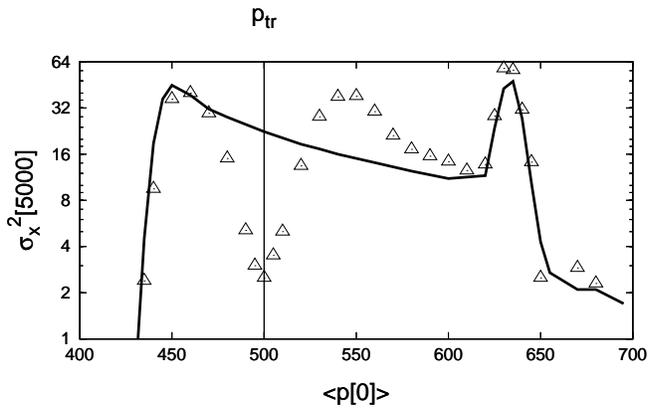}
\caption{The variance of atomic position $\sigma_x^2$ at
 $\tau=5000$ as a function of initial atomic momentum $\aver{p[0]}$:
curve --- constant field $\Delta(\tau)=-0.02$, triangles --- modulated field 
$\Delta(\tau)=-0.02+0.047\cos[0.00508\tau]$}
\label{fig11}
\end{center}
\end{figure}

At $0<|\Delta|\ll 1$ the atomic motion is more complex. The slowest atoms
($|\aver{p[0]}|<\sqrt{2m}$) are trapped in potential wells.
Faster atoms ($\sqrt{2m}\leq |\aver{p[0]}|< 2\sqrt{m}$) perform a kind
of random walk. Their wave packets split each time they cross standing-wave
 nodes, and this causes fast delocalization of the wave functions. The fastest atoms 
($|\aver{p[0]}|>2\sqrt{m}$) move ballistically through the wave. Their wave
 packets split, but all products move in the same direction, so the overall 
delocalization is slow. In Fig.~\ref{fig11} we calculate the variance 
of the atomic position $\sigma_x^2$ after a relatively long time span of coherent 
evolution $\tau=5000$ as a function of the initial
 atomic momentum $\aver{p[0]}$. For the constant field (solid curve) this function 
shows fast delocalization of all atoms in the range of 
$\sqrt{2m}\simeq 440\lesssim\aver{p[0]}\lesssim 2\sqrt{m}\simeq 640 $  
(cold atoms with velocities of the order of 1 m/s). Local peak at $\aver{p[0]}\simeq 630$
is produced by moderately fast atoms having an uncertain scenario of 
 either random walking or flying ballistically.

Now let us "switch on" the field modulation and see the changes. In 
Fig.~\ref{fig11} the analogous function of $\sigma_x^2$ is shown with triangles.
This function has a more complex structure. In particular, it has a prominent 
additional minimum at $\aver{p[0]}= p_{tr}\simeq 500$. These atoms 
are not trapped in potential wells in a strict sense (their
energy is too high, see the theory in the next sections), but
 some mechanism significantly suppresses the delocalization 
of their wave functions (note that both functions are shown in a logarithmic
scale).

\begin{figure*}[htb]
\includegraphics[width=0.96\textwidth,clip]{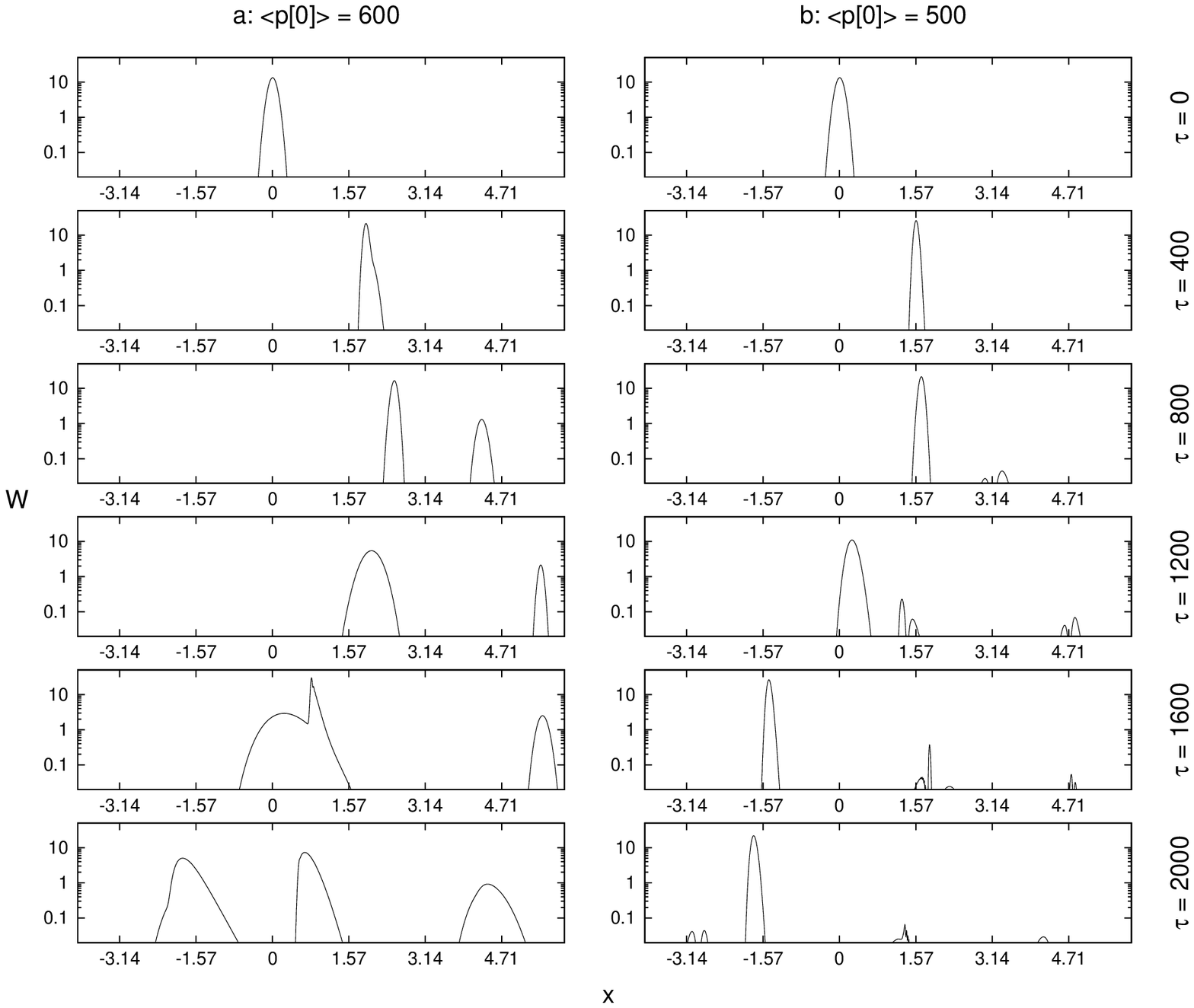}
\caption{Atomic wave packet splitting during quantum evolution (in the coordinate space):
(a) fast delocalization of typical wave function ($\aver{p[0]}=600$),
(b) slow delocalization of wave function in the velocity selective trapping mode ($\aver{p[0]}=500$).
Here $W[x]$ is the probability density to find an atom at coordinate $x$. Note:
due to logarithmic scale, in may seem that the norm of wavefunction is
not consered. However, our computations have shown that it is conserved
 with a good accuracy}
\label{fig22}
\end{figure*}

Let us consider the evolution of the corresponding wave packets in a coordinate
space. In Fig. \ref{fig22} we show the evolution of wave functions 
with $\aver{p[0]}=600$ and $500$ (other parameters are the same as in
 Fig.~\ref{fig11}). In both cases wave packets split. The first splitting occurs
 near the first node, $x\simeq 1.57$
(products overlap at $\tau=400$, but become completely independent at 
$\tau=800$). However, the proportion of splitting radically differs for 
$\aver{p[0]}=600$ and $500$. In Fig.~\ref{fig22}a fission products have
 similar "weights", while in Fig.~\ref{fig22}b 
they are radically different: a single large packet regularly oscillates in the
 range of $-2\lesssim x \lesssim 2$ "emitting" very small packets in both directions. 

We conclude that the slow delocalization of the wave function with 
$\aver{p[0]}\simeq 500$ is caused by the prominent asymmetry of
 wave packet splitting. Some mechanism suppresses the splitting 
of packets, and the atom is almost completely
 trapped in the range of $-2\lesssim x\lesssim 2$ (the variance
of its position $x$ is even smaller, see Fig.~\ref{fig11}). This suppression
is significant only for atoms with $490\lesssim \aver{p[0]}\lesssim 510$ 
(a comparatively narrow momentum and velocity range). 

\section{Explanation of the effect} 

In the previous section we used quantum equations to simulate atomic dynamics.
In this and the further section, in order to explain the effect of velocity selective trapping,
let us mention some semiclassical analytical results from \cite{pra,JETPL2009}	
(obtained for the stationary field). 

\begin{figure}[htb]
\begin{center}
\includegraphics[width=0.48\textwidth,clip]{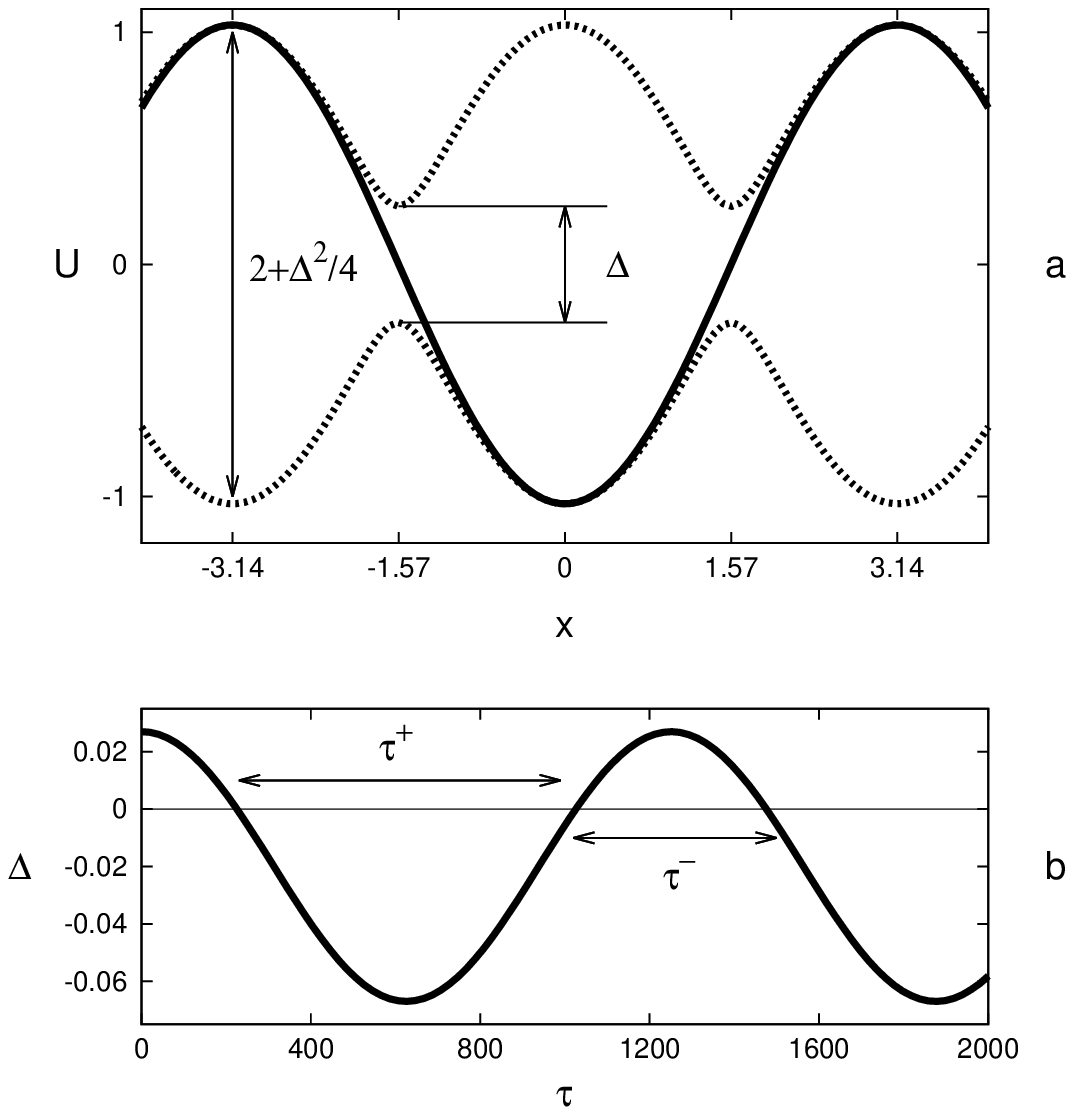}
\caption{(a) Periodic potentials in the space: dashed line --- non-resonant
 potentials $U^\pm$, solid line --- resonant potential $-\cos[x]$;
(b) illustration of the trapping condition: the modulation of detuning $\Delta[\tau]$ 
must be synchronized with atomic mechanical motion ($\Delta=0$ each time
a trapped wave packet crosses the standing wave node)}
\label{fig33}
\end{center}
\end{figure}

In a stationary field with $|\Delta|\ll 1$
 the atomic motion can be described in terms of two potentials
\be
U^-=- \sqrt{\cos^2[x]+\frac{\Delta^2}{4}},\quad U^+= \sqrt{\cos^2[x]+\frac{\Delta^2}{4}}.
\label{U}\ee                                    
(Fig.~\ref{fig33}a, dashed lines). An atom moves in one of these
potentials when it is far from the standing wave nodes. When an atom crosses
 the node, the potential may change
 the sign (atom undergoes the Landau-Zener tunneling between
 potentials $U^\pm$). The probability of tunneling depends on $\Delta$
as $\exp[-A\Delta^2]$ ($A$ is a combination of other parameters). At $0<|\Delta|\ll 1$ the tunneling causes splitting of
 the wave packet (observed in numerical experiments). At $\Delta=0$ potentials
coincide at nodes, so the probability of tunneling is equal to 1 and
wave packets do not split. The correspondent potential 
takes the simplest form $U=\pm \cos[x]$ (Fig.~\ref{fig33}a, solid line).

What happens, if we "switch on" the field modulation? When an atom
moves far from the nodes nothing radically changes. It moves in a 
constant potential that does not depend much on the value of $\Delta$.
Far from nodes we may neglect the term $\Delta^2/4$ in (\ref{U})
and put $U\simeq \pm \cos[x]$ with good accuracy. 

There are two possible scenarios when an atom crosses the node (at time $\tau$): (1) $\Delta[\tau]\ne 0$, therefore, the packet splits significantly;
(2) $\Delta[\tau] \simeq 0$, therefore, the splitting is suppressed. 
                        
The first scenario is more typical if the modulation is not synchronized with the atomic mechanical motion (because most of the time $\Delta[\tau]\ne 0$). Second
scenario may occur sometimes, but does not change the overall statistics
of the atomic motion. The evolution of the wave function shown in Fig.~\ref{fig22}a is
 typical for moderately small detunings $|\Delta|\sim 0.01$ (both
for the stationary and the modulated field). 

However, the evolution radically changes if the field
modulation is synchronized with the atomic mechanical motion. In particular,
 it is possible to choose such modulation parameters and atomic momentum
 (the particular values are estimated in the next section)
 that $\Delta[\tau]$ takes zero values  each time an atom crosses the node. 
With our parameters such synchronization occurs
 at $\aver{p[0]}= p_{tr}\simeq 500$ (Fig.~\ref{fig22}b).
Note that packet splittings are suppressed, but not completely. 
 Slight splittings are caused by the Landau-Zener
transitions that occur not exactly at a standing wave node, but in its small
vicinity (when $\Delta[\tau]$ is small but does not equal to zero).

\section{Estimation of trapping conditions}

Let us obtain the analytic relationship between trapping momentum $p_{tr}$ and
field parameters. Trapping occurs, if $\Delta[\tau]=0$ each time atom crosses
the nodes of the standing wave. In other moments of time $|\Delta[\tau]|\ll 1$.
Therefore, the term $\Delta^2/4$ in (\ref{U}) is always neglible, and
 the trapped atom moves in the effective potential
 $U\simeq-\cos[x]$ (we choose the negative sign of $U$, because in this paper
atoms with initial position $x[0]=0$ start their motion from the potential
 well). Therefore, the atomic center-of-mass  motion may be described
  by the semiclassical equations of motion \cite{PLA2011}
\be
\begin{aligned}
\dot x=&\ \frac{ p}{m},\quad
\dot p=-{\rm grad} [U]=-\sin[x],
\label{s}\end{aligned}\ee
with the trapping energy 
\be
E_{tr}\equiv\frac{ p^2}{2m}- \cos[x]=\frac{ p_{tr}^2}{2m}-\cos[\aver{x[0]}]
\label{E}
\ee
being the integral of motion (determined by the initial 
conditions). The trapping energy must be in the range of $0<E_{tr}<1$ (for $x[0]=0$,
this corresponds to $\sqrt{2m}<|p[0]|<2\sqrt{m}$). Slower atoms cannot reach
the standing wave node, and faster atoms move ballistically. 

Let us calculate the
 atomic traveling time between the two successive crossings of nodes in 
the negative and the positive segments of potential $-\cos[x]$ by integrating (\ref{s}) (for $0<E_{tr}<1$)
\begin{equation}
\begin{aligned} 
&\tau^-   =2k\sqrt{m}, 	 \quad k \equiv\sqrt{\frac{2}{1+E_{tr}}} ,	\\
& \tau^+=2k\sqrt{m}\left( F\left[\frac{\pi-|\arccos [E_{tr}]|}{2},k\right]-1\right),
\label{taupm}
\end{aligned} 
\end{equation}
Here $F$ is the elliptic integral of the first kind.

In order to synchronize the modulation with the atomic mechanical motion the time intervals 
$\tau^\pm$ must be equal to time
intervals between successive zeros of $\Delta[\tau]$ (Fig.~\ref{fig33}b).
Therefore, using (\ref{mod}) and (\ref{taupm}), we get
\begin{equation}
\zeta=\frac{2\pi}{\tau^-+\tau^+},\quad
\frac{\Delta_0}{\Delta_1}=-\cos\left[\frac{\pi\tau^-}{\tau^-+\tau^+}\right].
\label{fieldpa} 
\end{equation}

These formulae are true for atoms with
any initial positions (not only $x[0]=0$ used in (\ref{i})). 
At any value of atomic energy in the range of $0<E_{tr}<1$ (and appropriate
initial momentum) the velocity 
selective trapping of atoms can be achieved with appropriate values of 
$\Delta_{0,1}$, $\zeta$  calculated by these formulae. 
E.g., in order
   to observe trapping at $\aver{p[0]}=500$, $x[0]=0$,
 the field must have parameters $\zeta=0.00508$,
 $\Delta_0/\Delta_1=-0.4248$. We use them in numerical experiments,
 additionally fixing $\Delta_0=-0.02$.

\section{Conclusions}

In this paper we report the effect of velocity selective trapping of atoms in a
 frequency-modulated standing laser wave.

Intensive coherent light produces significant mechanical action on 
cold atoms having velocities of the order of $1$ m/s. There is a wide range
of field parameters at which atom performs a kind of random walk
accompanied with wave packets splitting and fast
delocalization of wave function. 
In this paper we report a specific field modulation mode that suppresses
wave packet splitting for atoms with precisely selected velocities.
These atoms oscillate in potential wells, and their wave functions are
almost completely localized.

This effect cannot cool atoms in the sense of achieving zero velocity, but it
can decrease their mechanical energy distribution. 
E.g., if we have a cloud of moderately cold
 atoms having wide position and momentum distribution we can switch on 
the modulated standing wave and wait for some time. Most of atoms
will leave the wave, and only small fraction will be trapped. These
trapped atoms will have similar mechanical energy determined by field parameters
 (see formulae (\ref{taupm}), (\ref{fieldpa})), and only the phase of their mechanical 
oscillations in wells will differ (because initial position distribution
 is random). In future study, we plan to simulate numerically
 large atomic ensemble cooled by the modulated laser. This will demonstrate
explicitly that our effect not only traps but also cools the atoms.

The effect of velocity selective trapping of atoms, being
theoretically predicted with the semiclassical apparatus, has been confirmed by
purely quantum numerical modeling. Therefore, it is not just an artifact
of semiclassical analytics but a real possibility. 
The drawback of this result is that it is obtained in absence of dissipation.
However, we believe that 
this is just a quantitative technical limitation that may be overcome by an
 appropriate choice of atoms and hi-Q cavities. 
                         
 This work has been supported by the Grant of the Russian Foundation for Basic Research
 12-02-31161.

\end{document}